\documentclass[a4paper]{jpconf}
\usepackage{graphicx}
\usepackage[load-configurations=abbreviations , separate-uncertainty , exponent-product = \cdot]{siunitx}

\newcommand{\jpsi}{\ensuremath{J\kern-0.05em /\kern-0.05em\psi} }
\newcommand{\sqrts}{\ensuremath{\sqrt{s}} } 
\newcommand{\pT}{\ensuremath{p_\mathrm{T}} }
\DeclareSIUnit\c{\textit{c}}
\DeclareSIUnit\MeVperc{\MeV\per\c}
\DeclareSIUnit\GeVc{\GeV/\c}
\DeclareSIUnit\TeVperc{\TeV\per\c}
\DeclareSIUnit\MeVpercc{\MeV\per\c\squared}
\DeclareSIUnit\GeVpercc{\GeV\per\c\squared}
\DeclareSIUnit\TeVpercc{\TeV\per\c\squared}

\begin{document}
\title{Measurement of \jpsi production in pp collisions at LHC energies with ALICE}

\author{S G Weber on behalf of the ALICE Collaboration}

\address{Research Division, GSI Helmholtzzentrum f\"ur Schwerionenforschung, Darmstadt, Germany}

\ead{S.Weber@gsi.de}

\begin{abstract}
An overview of ALICE results on the measurement of \jpsi production in pp collisions at \sqrts = \SI{7}{TeV} collected during the LHC Run-1 period is presented, as well as first results at forward rapidity from pp collisions at \sqrts = \SI{13}{TeV} collected during the LHC Run-2 period. In particular, the measurement of \jpsi production as a function of transverse momentum and charged-particle multiplicity are discussed and compared to theoretical model calculations.
\end{abstract}

\section{Introduction}

\subsection{Motivation}

The role of charmonium production as an excellent probe for the deconfined medium in ultra-relativistic heavy-ion collisions has been proposed a long time ago \cite{matsuisatz} and is universally accepted in the field. It is understood that a high quality baseline from proton-proton collisions is crucial for the quantification of hot medium effects taking place in nucleus-nucleus collisions, as well as cold nuclear matter effects in proton-nucleus collisions.

Besides that, hadronic charmonium production in proton-proton collisions is already a complex, intrinsically multi-scale process. In a simplified way it can be thought of as a two-stage process: in a first step charm-anticharm quark pairs are produced in elementary parton-parton interactions. This high energy scale process is well understood in the context of perturbative Quantum ChromoDynamics (pQCD). The subsequent binding into charmonium states happens at lower energy scales and is thus a non-perturbative process. There are mainly three theoretical frameworks that aim to describe hadronic charmonium production:

\begin{itemize}
\item Color Evaporation Model (CEM) \cite{cem}: It is based on quark-hadron duality. The cross section of a given charmonium state is proportional to the cross section of constituting heavy quark pair, independent of energy, transverse momentum or rapidity.
\item Color Singlet Model (CSM) \cite{csm}: pQCD is used to model the production of on-shell charm-anticharm quark pairs with the same quantum numbers as the charmonium states into which they will evolve, hence only color-singlet states are taken into account.
\item Non-Relativistic QCD (NRQCD) \cite{nrqcd}: Also contributions from color-octet states are considered. The transition into a color-singlet state is treated  non-perturbatively as an expansion in powers of the relative velocity between the two heavy quarks. It is parametrized using universal long-range matrix elements, obtained from fits to measured data sets.
\end{itemize}
These models make predictions on the transverse momentum and energy dependence of hadronic \jpsi production as well as the polarization.

\begin{figure}[t]
\begin{minipage}[t][][b]{15pc}
\includegraphics[width=15pc]{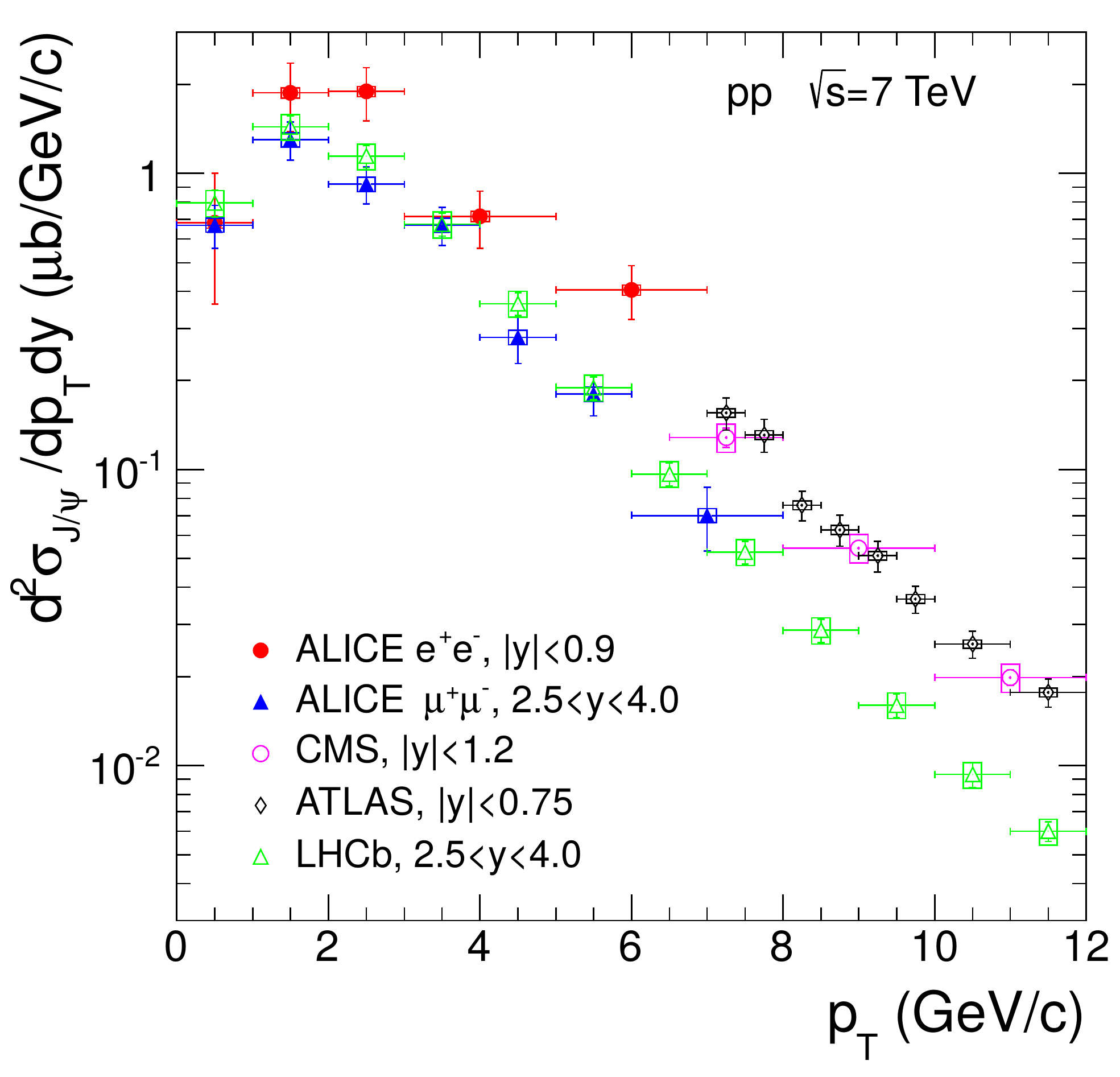}
\end{minipage}
\hspace{2pc}%
\begin{minipage}[t][][b]{20pc}
\includegraphics[width=20pc]{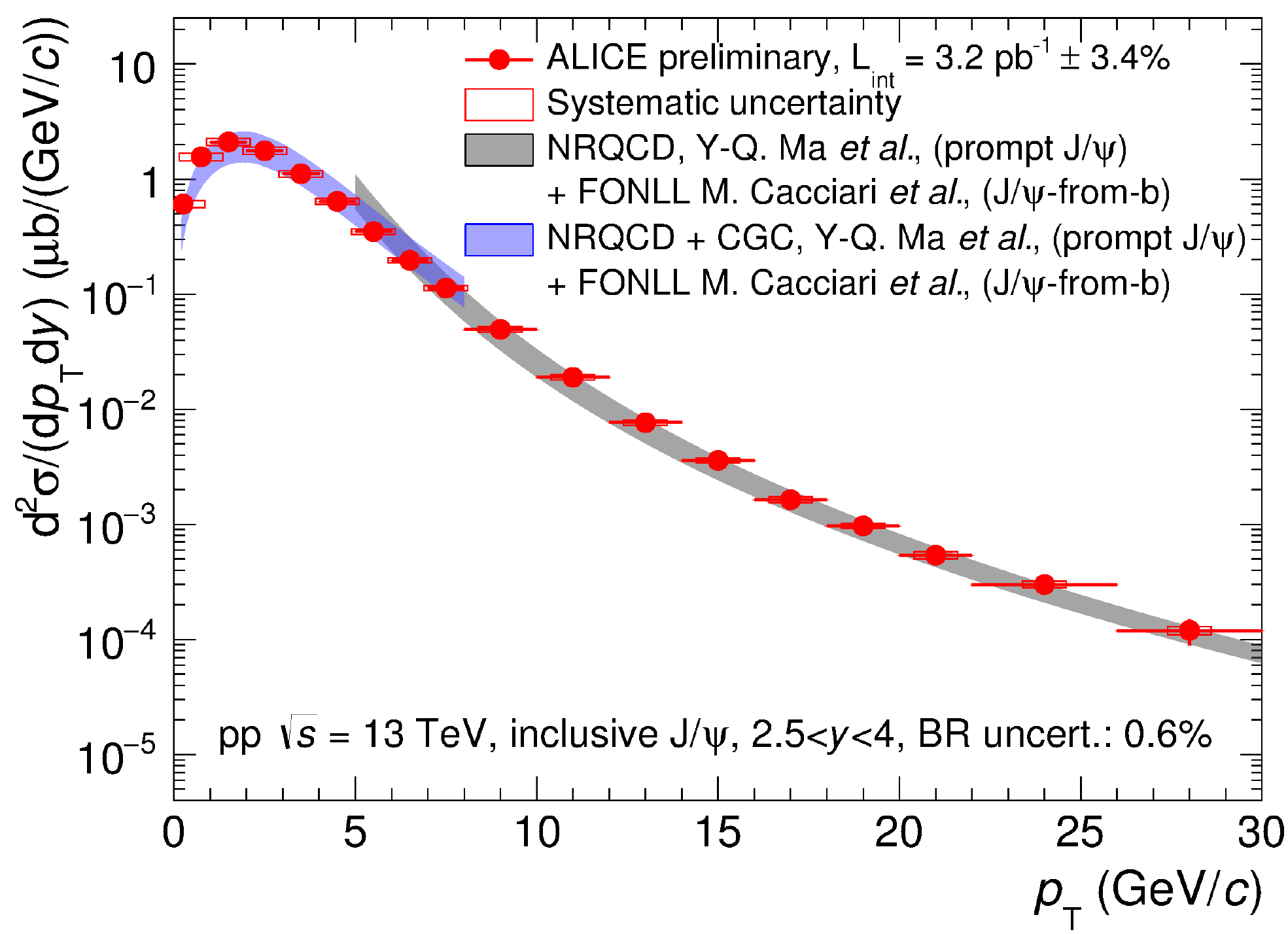}
\end{minipage} \caption{\label{pt} Inclusive \jpsi cross section in pp collisions as a function of transverse momentum. Left: Results at \sqrts = \SI{7}{TeV} at forward (blue triangles) and mid-rapidity (red squares). For comparison, results from other LHC experiments at the same collision energy are also shown. \cite[and references therein]{7tev, jpsi_fwd}. Right: Results at \sqrts = \SI{13}{TeV} at forward rapidity with model predictions from the NRQCD model \cite{nrqcd_13tev}, NRQCD and CGC \cite{nrqcd_cgc_13tev} for the prompt contribution and FONLL \cite{fonll} for the non-prompt contribution.}
\end{figure}

Another observable is the \jpsi yield as a function of the underlying event activity. It provides insight into the interplay between the hard and soft mechanisms relevant for charmonium production, and into the importance of multiple interactions \cite{sarah}. It might also be useful to address new proposed phenomena, such as the onset of collective effects in high multiplicity proton-proton collisions. In the analysis presented here, the event activity is quantified by the charged particle multiplicity at mid-rapidity.

\subsection{Experimental setup and analysis methods}

The ALICE detector is capable of reconstructing \jpsi particles in two kinematic regions in two different decay channels, i.e. at mid-rapidity  ($|y| < 0.9$) in the dielectron decay channel and at forward rapidity ($2.5 < y < 4.0$) in the dimuon decay channel. At mid-rapidity the Inner Tracking System is used for vertexing and tracking, and the Time Projection Chamber as the main tracking and Particle IDentification (PID) device. At forward rapidity, the dedicated muon arm is used for triggering, tracking and PID. The charged particle density is calculated using the number of tracklets reconstructed from hits in the two innermost layers of the Inner Tracking System. More details on the ALICE detector can be found in \cite{alice}, details on the analyses in \cite{7tev} and \cite{jpsi_mult}.
\newline
\newline
When comparing experimental results with model predictions one should keep in mind the different sources of \jpsi production: In an inclusive measurement about \SI{60}{\percent} of the \jpsi cross section is directly produced, \SIrange{20}{30}{\%} is feed-down from heavier charmonium states, and \SIrange{10}{20}{\%} comes from the weak decay of B-mesons. The first two sources are commonly referred to as prompt production, the last as non-prompt. Models often make predictions for direct or prompt \jpsi production, therefore for a quantitative comparison to data it is important to take all the components into account.

\newpage

\begin{figure}[t]
\begin{minipage}[t][][b]{15pc}
\includegraphics[width=15pc]{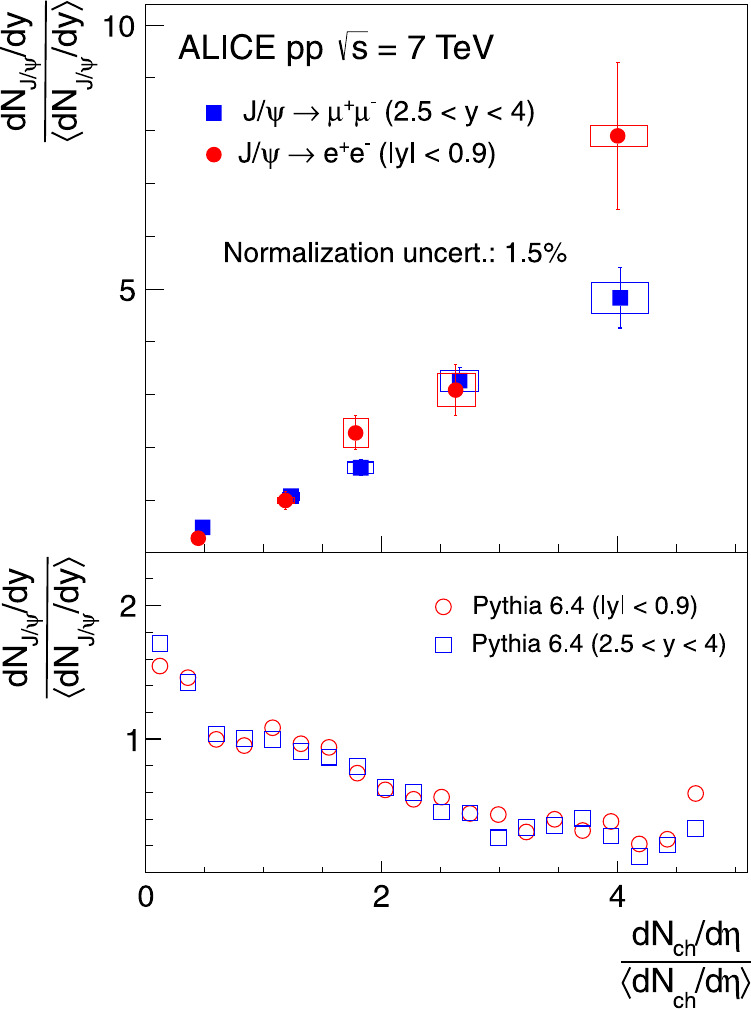}
\end{minipage}
\hspace{2pc}%
\begin{minipage}[t][][b]{20pc}
\includegraphics[width=20pc]{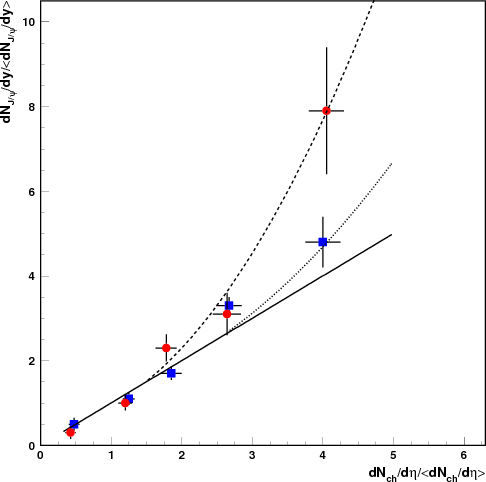}
\end{minipage}
\caption{\label{mult} \jpsi yield in pp collisions at \sqrts = \SI{7}{TeV} as function of charged particle multiplicity at mid-rapidity, both values being normalized to the corresponding mean values in minimum bias pp collisions. Blue squares refer to \jpsi at forward rapidity, red circles to \jpsi at mid-rapidity. Left: ALICE measurement (top), compared to prediction by PYTHIA6 (bottom) \cite{jpsi_mult}. Right: Prediction by percolation model \cite{perco}. The dotted line refers to \jpsi at forward rapidity, the dashed line to \jpsi at mid-rapidity, while the solid line shows the linear behavior.}  
\end{figure}

\section{Results} 

\subsection{Inclusive cross section as function of transverse momentum}

Figure \ref{pt} shows the inclusive \jpsi cross section in pp collisions measured by ALICE as a function of the transverse momentum. On the left, results at \sqrts = \SI{7}{TeV} at forward and mid-rapidity are shown, together with results from other LHC experiments. At forward rapidity the different LHC experiments are in good agreement with each other, at mid-rapidity the ALICE results complement the other experiments, since ALICE is the only experiment that can measure the cross section down to zero transverse momentum.
\newline
\newline
In Fig. \ref{pt} right, results at forward rapidity at \sqrts = \SI{13}{TeV} are shown, together with model predictions. In the transverse momentum range above \SI{5}{\GeVc} the prompt contribution to the \jpsi cross section is calculated in the NRQCD framework \cite{nrqcd_13tev}, below \SI{7}{\GeVc} the NRQCD model is combined with a description of the incoming protons in the Color Glass Condensate (CGC) model \cite{nrqcd_cgc_13tev}. The non-prompt contribution is calculated in the whole \pT range from the FONLL model \cite{fonll}. The different contributions are then summed to describe the inclusive production. The combination of these models provides a very good description of the measured cross section over the full transverse momentum range. In the overlapping region from \SIrange{5}{7}{\GeVc} the predictions from the NRQCD model and from the combined approach including CGC are in agreement with each other.

\newpage

\subsection{Self-normalized yields as function of self-normalized charged particle density}

Figure \ref{mult} shows the relative \jpsi yields at forward and at mid-rapidity as a function of the relative charged particle density around mid-rapidity, in pp collisions at \sqrts = \SI{7}{TeV}. On the top left plot the ALICE measurement is shown. In both rapidities an approximate linear increase can be seen, with the hint of a stronger than linear increase at mid-rapidity. On the left bottom plot, a prediction by PYTHIA 6.4.25 is shown, using the Perugia 2011 tune, taking into account only \jpsi directly produced in hard scatterings. The model predicts a decrease of the relative \jpsi yield with relative multiplicity, contrary to the observed increase. This difference can be interpreted as an indication for the relevance of other mechanisms for heavy-quark production, such as multi-parton interactions \cite{mpi}.

On the right side of figure \ref{mult}, a comparison of the data with a percolation model approach is shown. This model assumes a screening of particle multiplicities in regimes of high string densities. Together with the assumption of heavy quark production in initial collisions, a linear increase of relative production with multiplicity is expected at low relative multiplicities, changing into a quadratic behavior at high relative multiplicities, especially at mid-rapidity where higher densities are expected. The model is in good agreement with the data.

\section{Conclusions} 

ALICE measured the \pT differential inclusive \jpsi cross section at forward and mid-rapidity in pp collisions at various LHC energies down to zero transverse momentum. At mid-rapidity ALICE's capabilities at low transverse momentum are unique among the LHC experiments. The \pT dependent cross section at forward rapidity at \sqrts = \SI{13}{TeV} can be well described with a combination of predictions from the NRQCD model, a CGC description of the incoming protons at low \pT and FONLL predictions for the non-prompt contribution to the cross section. 

\jpsi production in pp collisions at \sqrts = \SI{7}{TeV} rises linearly with the mid-rapidity charged particle density, both at forward and mid-rapidity, with a hint of a stronger than linear increase at mid-rapidity. This trend hints to an important role of other mechanisms in charmonium production beside hard initial scattering processes, such as multi-parton interactions. The results are also in agreement with the percolation scenario.

\section*{References}
\bibliographystyle{iopart-num}
\bibliography{literatur}

\end{document}